\documentclass[twocolumn,prl,superscriptaddress,showpacs]{revtex4}

\usepackage{amssymb}
\usepackage[dvips,final]{graphicx}
\usepackage{dcolumn}
\usepackage{bm}
\begin{document}

\title{$^{59}$Co NMR study of the Co states in superconducting and anhydrous
cobaltates}

\author{I.R.~Mukhamedshin}
\altaffiliation[Permanent address: ]{%
Physics Department, Kazan State University, 420008 Kazan, Russia
}%
\affiliation{%
Laboratoire de Physique des Solides, UMR 8502, Universit\'e Paris-Sud, 91405
Orsay, France
}%

\author{H.~Alloul}%
\email{alloul@lps.u-psud.fr}
\affiliation{%
Laboratoire de Physique des Solides, UMR 8502, Universit\'e Paris-Sud, 91405
Orsay, France
}%

\author{G.~Collin}
\affiliation{ Laboratoire L\'eon Brillouin, CE Saclay, CEA-CNRS, 91191
Gif-sur-Yvette, France
}%

\author{N.~Blanchard}%
\affiliation{%
Laboratoire de Physique des Solides, UMR 8502, Universit\'e Paris-Sud, 91405
Orsay, France
}%

\begin{abstract}
$^{59}$Co NMR spectra in oriented powders of the superconducting (HSC)
Na$_{0.35}$CoO$_{2}$,1.3H$_{2}$O and Na$_{0.35}$CoO$_{2}$ compounds reveal a
single electronic Co state with identical $T$ independent NMR shift tensor.
These phases differ markedly from Na$_{0.7}$CoO$_{2}$, in which we resolve 3
types of Co sites. The large T variation of their spin susceptibilities
$\chi_{s}$ and the anisotropy of the orbital susceptibility $\chi_{orb}$ allow
us to conclude that charge disproportionation occurs, in a non magnetic
Co$^{3+}$ and two magnetic sites with about 0.3 and 0.7 holes in the $t_{2g}$
multiplet. The data are consistent with those for the single Co site in the
anhydrous and HSC phase assuming the expected Co$^{3.65+}$ charge.
\end{abstract}


\maketitle

The superconductivity (SC) induced by water insertion in Na$_{x}$CoO$_{2}$
\cite{Takada} has led to suggest an analogy with High $T_{c}$ cuprates \cite
{Baskaran}. Indeed, metallic and magnetic states \cite{Sugiyama} occur in the
phase diagram of these layered transition metal oxides and the SC, which occurs
for a limited range of carrier contents \cite{Cava}, might be explained by the
magnetic scenarios proposed for the cuprates. The triangular lattice of the
cobalt planes introduces a frustration of interactions between Co spins which
could be even more favorable than in the cuprates to the formation of exotic
spin ground states \cite{Kumar}. Furthermore, cobaltates might display a richer
variety of properties due to the two possible Co electronic configurations
(Co$^{3+}$, $S=0$ and Co$^{4+}$, $S=\frac{1}{2}$) in the high trigonal crystal
field of the CoO$_{2}$ plane. A segregation into these $3^{+}$ and $4^{+}$
charges has been suggested \cite{Ray} and results, at least for $x_{0}\approx
0.7$, in a Co charge order correlated with the atomic arrangement of the Na
layers \cite{NaPaper}. This charge ordered metallic state, presumably
responsible for the thermoelectric properties \cite{Terasaki}, should be
physically quite distinct from the insulating charge and spin stripe order due
to hole localization in some cuprates \cite{Tranquada}.

Prior to fully addressing these comparisons, it seems essential to perform
atomic probe determination of the Co states all over the phase diagram. This
issue is connected with that of the role of H$_{2}$O in inducing SC. While
H$_{2}$O insertion should not \textit{a priori} modify the electronic structure
\cite{Johannes}, it has been recently proposed \cite{Takada2,Mylne} that
(H$_{3}$O)$^{+}$ ions increase the doping of the CoO$_{2}$ plane from 0.35 to
0.7, so that the electronic properties of the SC would be similar to those of
the $x_{0}\approx 0.7$ phase.

In the present work we use $^{59}$Co NMR, in oriented powders in order to probe
the electronic properties of the Co sites. The analysis of the NMR shifts
enables us to determine the orbital susceptibility $\chi^{orb}$ which is
directly related to the charge state of the Co. We show for the first time
\cite{Imai} that three types of Co states occur in the $x_{0}$ phase at
variance with the anticipated Co$^{3+}$/Co$^{4+}$ scenario \cite{Lee}. On the
contrary, the $x\approx 0.35$ sample is found to display a uniform Co$^{3.65+}$
charge state which is not modified upon water insertion. We therefore stress
the contrast between the electronic properties on the two sides of the $x=0.5$
composition. The large anisotropy of $\chi^{orb}$ found for the hole doped Co
sites allows us to propose a scheme to reconcile the conflicting data
\cite{Kobayashi,Waki} reported below T$_{c}$ in the HSC samples, and to propose
singlet SC. Our results open the way for experimental electronic
characterization of the various magnetic states found for large Na content
\cite{Mendels}, and should stimulate realistic band structure calculations.

Two $x_{0}\approx 0.7$ and the $x\approx 0.35$ samples were already used in
Ref.~\onlinecite{NaPaper}. Their NMR signals did not evolve during the last
18 months. A new $x_{0}$ sample with negligible Co$_{3}$O$_{4}$ content has
been synthetised. A part of the $x\approx 0.35$ batch was hydrated into a
HSC sample ($T_{c}$=$3.9$~K) and the $c$ axes of the sample crystallites
were aligned in water in a 7~T applied field. The sample was then frozen and
kept below 0${^{\circ }}$C.

\begin{figure}[tb]
\center
\includegraphics[width=1\linewidth]{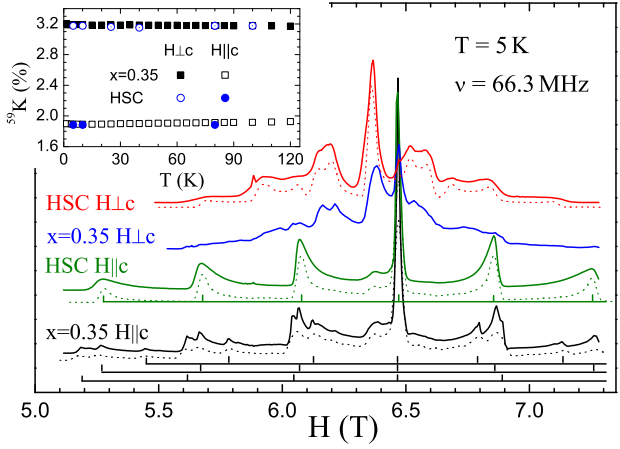}
\caption{$^{59}$Co NMR spectra in the $x\approx 0.35$ samples. Dotted lines are
simulations with $\nu_{Q}$=4.1~MHz and $\eta$=0.22 (HSC sample) and $\nu_{Q}=
$3.55, 4.17 and 4.45~MHz with intensity ratios 47/30/23 (anhydrous sample).
Insert: $T$ variation of the $^{59}$Co NMR shifts.}
\label{fig:1}
\end{figure}

The $I$=$7/2$ nuclear spin of $^{59}$Co both senses the magnetic properties of
the Co site and couples through its nuclear quadrupole moment $Q$ to the
electric field gradient (EFG) tensor $V_{\alpha\beta }$ created by its charge
environment. The $^{59}$Co NMR signal detected for $H\Vert c$ in the HSC sample
is the most typical spectrum for a single site for which $c$ is principal axis
of the EFG. It is reported in Fig.~\ref{fig:1} as a recording
versus $H$ of the NMR spin echo intensity obtained with a $\frac{\pi }{2}%
-\tau -\frac{\pi }{2}$ pulse sequence. It displays a set of seven equally
spaced peaks symmetrically arranged around the central $-\frac{1}{2}%
\leftrightarrow \frac{1}{2}$ transition. The frequency splitting $\nu
_{Q}=3eQV_{ZZ}/(h2I(2I-1))$ is a measure of the EFG component $V_{ZZ}$ with $%
Z\Vert c$. For $H\perp c$ the in plane $ab$ axes of the crystallites are random
in our samples, which results in a powder spectrum. The lineshape reflects then
the anisotropy $\eta =\left| (V_{YY}-V_{XX})/V_{ZZ}\right| $
of the in plane components of the EFG. The simulations of the spectra yield $%
\nu _{Q}$=4.103(10)~MHz; $\eta $=0.22(2), in perfect agreement with NQR
existing data \cite{Kitaoka}. The spectra for the anhydrous $x\approx 0.35$
sample are quite similar to that of the HSC sample. The small difference
relates to slight discrete changes of $\nu _{Q}$ resolved only for $H\Vert c$%
. These can easily be attributed to EFG contributions of diverse local
arrangements of Na$^{+}$ ions which are negligible in the SC samples due to
the larger Na$^{+}$ - Co distances upon water insertion. Remarkably, the
spectra in the anhydrous sample display a unique central line for all Co
sites. Furthermore, its anisotropic NMR shifts, which measures the local $%
\chi$, are identical and nearly $T$ independent in the two samples ($K_{c}$=%
$1.9\%$ and $K_{ab}$=$3.2\%$, see Fig.~\ref{fig:1} and insert) \cite{Shift}.
This is strong evidence that \textit{the HSC and anhydrous samples display a
quite identical uniformly charged} \textit{state, e.g. }Co$^{3.65+}$\textit{%
, with no resolvable difference in local }$\chi$\textit{, whatever the Na
environment.}

\begin{figure}[tb]
\center
\includegraphics[width=1\linewidth]{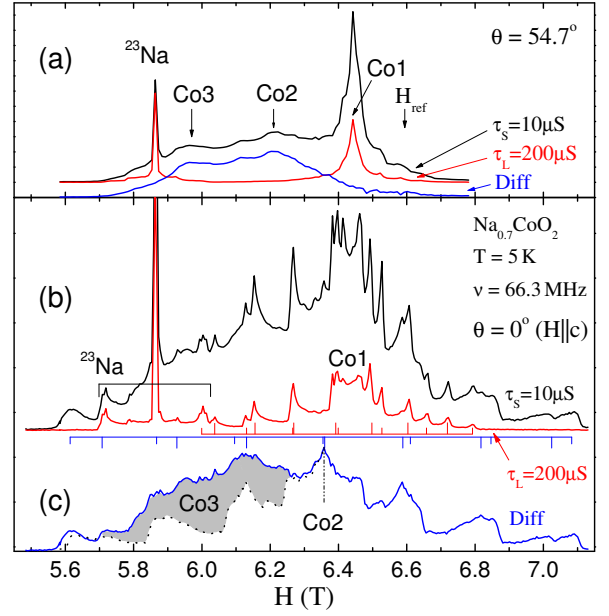}
\caption{$^{59}$Co NMR spectra taken as the $H$ dependence of the spin echo
taken for short ($\tau_{S}$) or long ($\tau_{L}$) delay between pulses.
\textit{Upper part} (a) for $H$ tilted at the ''magic angle'' $\theta $.
\textit{Lower part} for $H\Vert c$: (b) the ($\tau_{L}$) spectrum corresponds
to the Co1a and Co1b sites. (c) The difference reveals the short $T_{2}$ sites
Co2a Co2b and Co3. The quadrupole satellites of the Co1 and Co2 sites are
pointed by ticks.}
\label{fig:2}
\end{figure}

The situation is opposite for the spectra of the $x_{0}\approx 0.7$ phase
which display many different Co sites (Fig.~\ref{fig:2}b) with distinct EFGs
and magnetic shifts. To resolve the latter we took spectra with $H$ tilted
of $\theta =(\widehat{H,c})$, such that $3cos^{2}\theta =1$. For that
''magic angle'' (MA) the quadrupole satellites of the lines with $\eta =0$
merge with their central lines and the quadrupole spectra for $\eta \neq 0$
are also narrowed (Fig.~\ref{fig:2}a). Besides the $^{23}$Na signal one can
easily distinguish then three Co sites, indexed with increasing shifts Co1,
Co2 and Co3. The Co2 line has the largest broadening, to be associated
hereafter with a large anisotropy of its NMR shift. In Fig.~\ref{fig:2}a we
also show how we contrast the signals of these sites using their very
distinct transverse spin spin relaxation time $T_{2}$. Indeed, a spin echo
signal dies away for $\tau\gg T_{2}$, and for a short delay time ($\tau_{S}$%
) the MA spectrum displays all sites, while for a long delay time ($\tau_{L}$%
) only the Co1 and Na signals remain. This reveals that the signals of the
''more magnetic'' Co2 and Co3 sites have $T_{2}<\tau_{L}$. Once the Co1
spectrum is determined with the ($\tau_{L}$) data, it can be rescaled and
deduced from the ($\tau_{S}$) spectrum to isolate the Co2 and Co3 signals
(''Diff'' in Fig.~\ref{fig:2}).

This procedure has been quite powerful in resolving the spectra for any $%
\theta $ as exemplified for $H\Vert c$ in the lower part of Fig.~\ref{fig:2}%
. The ($\tau _{L}$) signal for Co1 is seen to display more than 7 well
resolved quadrupolar lines and a full analysis reveals that it consists of
the superposition of spectra of two sites Co1a and Co1b with similar shifts
and slightly different  $\nu _{Q}$. The spectrum of Co2 and Co3 in Fig.~\ref
{fig:2}c reveals that Co2 consists also of a set of quadrupole lines which
also correspond to two slightly different sites Co2a and Co2b with $\nu _{Q}$
about twice larger than for Co1. The highly shifted Co3 component appears on
top of the simulated Co2 spectrum. In this Letter, the lower part of Fig.~%
\ref{fig:2} is displayed to let the reader understand the method used, but
the parameters listed in Table~\ref{tab:table1} result from information
taken on more than a hundred spectra at different $T$ and $H$. We postpone
the details on the EFG data to a full technical report and focus hereafter
on the physical information extracted from the shifts.

\begin{figure}[tb]
\center
\includegraphics[width=1\linewidth]{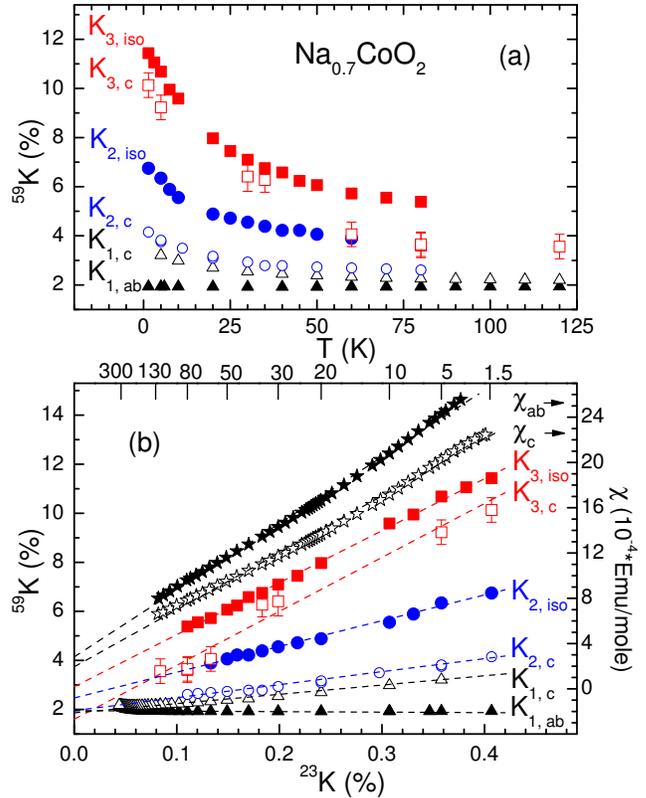}
\caption{(a) $T$ variation of the shifts $K_{i,\alpha}$ of the Co sites.
$K_{i,iso}$ are taken from MA spectra. (b) NMR shift data for the various Co
sites (left scale) and SQUID data for $\chi_{\alpha}^{m}$ (right scale) are
plotted versus the $^{23}$Na NMR shift (Fig.~4 in Ref.~\onlinecite{NaPaper}).}
\label{fig:3}
\end{figure}

The large $T$ variations of of $^{59}$Co NMR shifts, which reflect the
anomalous magnetism of the CoO$_{2}$ planes \cite{NaPaper}, are detected in all
cases except for the Co1 sites for $H\perp c$, as reported in Fig.~\ref
{fig:3}a. For this reason this signal had been attributed to non magnetic Co$%
^{3+}$ in Ref.~\onlinecite{Ray}. We do find here that for $H\Vert c$ the Co1
and Co2 sites exhibit a similar increase of $K_{c}$ with decreasing $T$, and a
larger one for Co3, although less accurately determined. For $H\perp c,$it is
harder to resolve the powder spectrum of Co2 similar to that of Fig.~\ref
{fig:1}. We better monitored the anisotropy of the Co2 and Co3 shifts by
comparing $K_{c}$ to the isotropic $K_{iso}=\sum K_{\alpha}/3$ obtained from MA
spectra.

\begin{table}[b]
\caption{NMR parameters for the 3 distinct sites and subsites for the samples
with $x_{0}\approx 0.7$.}
\label{tab:table1}%
\begin{ruledtabular}
\begin{tabular}{lcccccc}
Site & ${I_{i}\footnotemark[1]}$ & ${K_{iso}^{orb}}$ & ${K_{iso}^{s}/^{23}K}$ &
${K_{c}^{orb}}$ & ${K_{c}^{s}/^{23}K}$ & $\nu_{Q}$ \\
& (\%) & (\%) & & (\%) & & (MHz)\\
\hline $\left.
\begin{tabular}{l}
Co1a \\
Co1b \\
\end{tabular}
\right\}$ & 26(4) &
\begin{tabular}{c}
1.89(2)\\
1.89(2)\\
\end{tabular}
&\begin{tabular}{c}
2.58(7)\\
2.16(7)\\
\end{tabular}
&\begin{tabular}{c}
1.95(1)\\
1.95(1)\\
\end{tabular}
&\begin{tabular}{c}
4.0(1)\\
3.3(1)\\
\end{tabular}
&\begin{tabular}{c}
1.19(1) \\
1.38(1) \\
\end{tabular}
\\
$\left.
\begin{tabular}{l}
Co2a \\
Co2b \\
\end{tabular}
\right\}$ & 55(5) & 2.5(1) & 10.5(3) &
\begin{tabular}{c}
1.92(5)\\
1.95(4)\\
\end{tabular}
&\begin{tabular}{c}
4.6(2)\\
5.3(2)\\
\end{tabular}
&\begin{tabular}{c}
2.18(2)\\
2.55(2)\\
\end{tabular}
\\
$\,\,$Co3 & 19(4) & 2.9(1) & 21.2(2) & 1.6(4) & 22(2) & $\lesssim$1.6 \\
\end{tabular}
\end{ruledtabular}
\footnotetext[1]{The relative occupancies $I_{i}$ of the Co1 and Co2 obtained
on $H\Vert c$ spectra, from the outer well resolved $\frac{5}{2}\rightarrow
\frac{7}{2}$ satellite intensities are $I_{1a}/I_{1b}$=1.9(2) and
$I_{2a}/I_{2b}$=0.5(1). All $^{59}$Co nuclei being detected in the anhydrous
$x\approx 0.35$ sample, its central line intensity $I_{(0.35)}$ was used to
calibrate the Co1 central line intensity. The ratio $I_{1}/I_{(0.35)}$=0.22(3)
found ensures that all Co are detected as well in the $x_{0}$ phase within
experimental accuracy.}
\end{table}

The macroscopic susceptibility taken in a direction $\alpha$%
\[
\chi_{\alpha}^{m}=\chi^{dia}+\chi_{\alpha}^{orb}+\chi_{\alpha}^{s}=\chi^{dia}+\sum\nolimits_{i}\left[
\chi_{i,\alpha}^{orb}+\chi_{i,\alpha}^{s}(T)\right]
\]
involves the diamagnetism of the ion cores and the orbital and spin terms
splitted in their Co($i$) sites components. The NMR shifts of sites $i$ are
linked to the $\chi^{\prime }s$ through specific hyperfine couplings $A$
\[
K_{i,\alpha}=K_{i,\alpha}^{s}+K_{i,\alpha}^{orb}=\;A_{i,\alpha}^{s}\;\chi_{i,\alpha}^{s}(T)\,+\,A_{i,\alpha}^{orb}\;\chi_{i,\alpha}^{orb}.
\]
The $^{59}$Co diamagnetic shift contribution is included in the reference taken
\cite{Shift}. As the $^{23}$Na shift \cite{NaPaper} is purely due to the $T$
dependent spin term, it is essential to compare $^{59}K_{i,\alpha}$ to
$^{23}K_{iso}$. As can be seen in Fig.~\ref{fig:3}b a linear dependence is
found for all sites below 120~K when the Na motion is frozen. So these data
confirm that all Co pertain to the same phase in which a single $T$ variation
characterizes the local $\chi_{i,\alpha}^{s}(T)$. The slopes of the linear fits
give the relative magnitude of the spin contributions
$K_{i,\alpha}^{s}/^{23}K$, while their $^{23}K$=0 intercept give the estimates
of the Co orbital NMR shifts $K_{i,\alpha}^{orb}$ reported in
Table~\ref{tab:table1}. For comparison SQUID data for $\chi_{\alpha}^{m}(T)$
taken in $H$=5~T for our oriented sample with minimal Co$_{3}$O$_{4}$ content,
also plotted in Fig.~\ref{fig:3}b versus $^{23}K,$ confirm this unique $T$
dependence. These data allow us to separate $\chi_{\alpha}^{s}(T)$ from
($\chi^{dia}+\chi_{\alpha}^{orb})$, the latter being the extrapolation to
$^{23}K$=0. With tabulated values we calculate for Na$_{0.7} $CoO$_{2}$
$\chi^{dia}=-0.35~$ and then $\chi_{ab}^{orb}=3.1(2)~$ and
$\chi_{c}^{orb}$=$2.3(2)$, in $10^{-4}$ emu/mole units \cite{Susc}.

This anisotropy of orbital $\chi$, seen on both the macroscopic
$\chi_{\alpha}^{orb}$ and the local $K_{i,\alpha}^{orb}$, being mainly of ionic
origin, give us an insight on the ionic states of the Co sites. For a Co$%
^{3+}$ ionic state, the lower energy $t_{2g}$ triplet levels are filled and the
ionic shell has a spherical symmetry, so that $\chi^{orb}=2\mu _{B}^{2}/\Delta
$ is isotropic. Here $\Delta $ is the $t_{2g}-e_{g}$ energy
splitting. Holes introduced on the $t_{2g}$ levels yield modifications of $%
\chi_{i,\alpha}^{orb}$ which depend of the spatial direction of the populated
hole orbitals.Therefore the isotropy of $K_{1,\alpha}^{orb}$ allows us to
assign Co$^{3+}$ to the Co1 site, while the anisotropic values of
$K_{2,\alpha}^{orb}$ and $K_{3,\alpha}^{orb}$ imply that Co2 and Co3 have hole
contents $x_{i}$, which will be estimated from our data. A first relation
between the $x_{i}$ is given, for $x_{0}\approx 0.7$, by the charge neutrality
condition $0.55x_{2}+0.19x_{3}=0.3,$ using the fractional occupancies $I_{i}$
of the Co sites given in Table~\ref{tab:table1}. With the rough but sensible
assumption that the orbital anisotropy scales with the hole content $x_{i}$, we
obtain $x_{2}$/$x_{3}\approx 0.44$ from the data for
$K_{i,iso}^{orb}$-$K_{i,c}^{orb}$ listed in Table~\ref{tab:table1}, which
results in charge states Co$^{+3.3(1)}$ and Co$^{+3.7(1)}$ respectively for Co2
and Co3.

In the light of these results we may now reexamine the anisotropic $^{59}$Co
shift data found hereabove in the $x\approx 0.35$ samples. As the 0.65 hole
content of the $t_{2g}$ orbital in the latter samples is similar to that
estimated for the Co3 site we might expect similar orbital shifts
$K_{c}=1.6(4)\%,$ $K_{iso}^{orb}\approx 2.9(1)\%$, that is $K_{ab}^{orb}\approx
3.5(4)\%$. These values are remarkably close to the total shifts $K_{c}=1.9\%$,
$K_{ab}=3.2\%$, reported in Fig.~\ref{fig:1} in the HSC sample, in which case
the spin shifts $K_{\alpha}^{s}$ are then quite small in the HSC samples. In
case of singlet SC these spin shifts are expected to decrease below $T_{c}$ and
to vanish at $T$=0. Two conflicting reports indicate (i) a decrease from
$K=3.2$ to $2.7\%$ \cite{Kobayashi} suggesting singlet SC and (ii) no variation
of $K=1.9\%$ \cite{Waki} pointing towards triplet SC. Comparing their normal
state data with ours, it is clear that the authors of (i) and (ii) respectively
monitored the $H\perp c$ and the $H\Vert c$ central line singularities in
partially oriented samples. So their results do not contradict each other if
the spin shifts are $K_{ab}^{s}\approx 0.5\%$ and $K_{c}^{s}\approx 0$. In that
case the data of (i) strongly supports spin singlet SC. We can further notice
that  $K_{c}^{s}\approx 0$  would be consistent with our observation (see
Table~\ref{tab:table1}) that $K_{i,c}^{orb}\approx 1.9\%$ is nearly independent
of the charge state of the Co. This therefore leads us to suggest that
\textit{the data are quite consistent and strongly support singlet SC}. To
conclude on the properties of the HSC phase, we have demonstrated here that
water insertion does not modify the electronic properties for $x\approx 0.35$,
which contradicts recent proposals \cite{Takada2,Mylne}. Therefore, the absence
of SC in the anhydrous sample could result from a depression of $T_{c}$ induced
by the Na$^{+}$ ion disorder. Even if such a disorder also exists in the HSC
samples one can anticipate that it might be screened then by the water layers,
and would less affect the CoO$_{2}$ plane carriers.

As for the $x_{0}\approx 0.70$ sample we have demonstrated that the fraction of
Co$^{3+}$ is quite small, which totally discards the purely ionic Co$^{3+}
$/Co$^{4+}$ picture, which would presumably yield a Mott Hubbard insulating
state. Our results allow us to establish that extracting Na$^{+}$ from
NaCoO$_{2}$ rather leads to delocalized states on specific Co sites, while only
a few Co sites remain Co$^{3+}$. This evidence for partial hole filling on Co2
and Co3 sites is more compatible with a metallic state, such as a commensurate
CDW state, although a nearly localized behaviour is still required to explain
the drastic $T$ variation of the local $\chi$. We have emphasized here the
different magnetic characters of the Co sites through the modifications of
$K_{\alpha}^{orb}$. The spin shifts $K_{i,\alpha}^{s}$
do also probe the local magnetism and the fact that $%
K_{i,iso}^{s}/^{23}K_{iso}$ is small ($\approx $2.3) for Co1 and increases to
$\approx $10 and $\approx $20 for Co2 and Co3, is in perfect qualitative
agreement with an increase of hole content on the $t_{2g}$ multiplet for these
sites. However a nuclear spin is coupled both to its own electron orbitals
giving an on-site hyperfine field (HF) and to its nearest neighbours
(transferred HF). So, further analysis of the $K_{i,\alpha}^{s}$ data require a
knowledge of the arrangement of the differently charged Co sites to undertake
electronic structure estimates of the hyperfine couplings $A_{i,\alpha}^{s}$.
Although the EFG's and the site occupancies will help in this process, special
care is presently devoted to secure first the atomic structure through x rays
and neutron investigations, since it is clear that the atomic arrangement of Na
is correlated to that of the Co charges.

It is often soundly suggested that the Na1 which sits on top of the Co atoms
stabilize the less positively charged Co$^{3+}$, so that the role of the Na
order could be essential for the physics of the $x\geq 0.5$ phases. The fact
that large high $T$ Curie-Weiss dependencies occur quite generally for $%
x>0.5$ might as well indicate that magnetism is an incipient property of the
CoO$_{2}$ planes with sufficient hole doping on the Co. The subtle Na ordering
could also play a second, more marginal 3D role, by favouring $c$ axis
couplings between hole doped magnetic Co ions, which might explain the diverse
low $T$ ordered magnetic states \cite{Mendels} found for  $x>0.75.$ We believe
that the methods used here to determine the Co charge states can be applied for
other Na ordered phases and should help to resolve these issues. In any case
the results of the present work strongly suggest that magnetic models should
not consider all Co sites on the same footing and will not be simple
realizations of Heisenberg spin Hamiltonians. Finally, it is of course obvious
at this stage that our mainly ionic approach can only be qualitative, as we are
dealing with an original organization of Co charges in a strongly correlated
band in which carriers avoid some sites. Further progress will have to await
realistic band structure calculations taking into account the correlations,
which could reproduce both charge disproportionation and anomalous magnetism.

We would like to acknowledge J.~Bobroff and P.~Mendels for their help on the
experimental setup, for stimulating discussions and for their careful reading
of the manuscript.


\begin{references}
\bibitem{Susc} These extrapolated SQUID data exceed of about 10$^{-4}$emu/mole
the values $\chi_{ab}^{orb}$=1.79 and $\chi_{c}^{orb}$=1.25
(10$^{-4}$emu/mole), which can be estimated from the data for
K$_{i,\alpha}^{orb}$ and $I_{i}$ of Table~\ref{tab:table1}, assuming a single
value for $A_{i,\alpha}^{orb}$ whatever the $3d$ orbital i.e. $A^{orb}$=
2$<r^{-3}>$=13.4 a.u, from Ref.~\onlinecite{Ray}. Experimental inaccuracies in
the extrapolations and the rough above assumption might explain this 50\%
difference.
\end{references}
\end{document}